\documentstyle[12pt]{article}

\newcommand{\ds}{\displaystyle}
\newcommand{\ltwid}{\raise.3ex\hbox{$<$\kern-.75em\lower1ex\hbox{$\sim$}}}
\newcommand{\rtwid}{\raise.3ex\hbox{$>$\kern-.75em\lower1ex\hbox{$\sim$}}}

\title{The Coleman-Weinberg effective potential \\
       in the theory of superconductivity}
\author{{\sl Rachel~M.~Quick$^{\ast}$
        and Sergei~G.~Sharapov$^\dagger$}\\
{\sl Department of Physics, University of Pretoria,}\\
{\sl 0002 Pretoria, South Africa}}

\date{}

\begin{document}

\maketitle


\begin{abstract}
A quasi two-dimensional non-relativistic four-Fermi theory is studied
at finite temperatures in the next-to-leading order approximation
using the Coleman-Weinberg effective potential. The appearance of an
imaginary part to the one-loop correction is discussed in the context of
condensed matter theory where it is referred to as the Thouless criterion
for superconductivity.  By reference to the appropriate modified effective
potential one may revise the Thouless criterion to obtain a critical
temperature in next-to-leading order that, unlike the mean-field
temperature, tends to zero in the two-dimensional limit in agreement with
the Coleman theorem.
\end{abstract}



\vfill

\noindent
{\sl $\dagger$ On leave of absence from Bogolyubov Institute for
Theoretical Physics of \\
the National Academy of Sciences of Ukraine, 252143 Kiev, Ukraine}

\noindent
{\sl $\ast$ {\em Corresponding author:} R.M.~Quick \\
Department of Physics, University of Pretoria,\\
Pretoria 0002 SOUTH AFRICA \\
E-mail: rcarter@scientia.up.ac.za}

\eject

\section{Introduction}

Low dimensional quantum field theories have recently attracted
a great deal of attention. This is in part because the development
of the theory of superconductivity applicable to
high-temperature superconductors (HTSC) \cite{Bednorz}
demonstrated the need for theoretical methods
which go beyond the standard Bardeen-Cooper-Schrieffer
(BCS) \cite{Schrieffer} mean-field approximation. Within the BCS
theory the fluctuations, or in other words next-to-leading order
corrections are regarded as a small correction to the mean-field
(leading-order) results. As such they do not lead to significant
renormalization of such important characteristics as the critical
temperature, $T_c$, of the superconducting transition.

By contrast, the influence of fluctuations in nonconventional
superconductors may result in the significant decrease
of $T_c$. The physical reason for this is rather simple. HTSC
have, in contrast to conventional superconductors, a lower
dimensionality of space for carrier motion and a smaller carrier
density. It is known that these factors increase the influence of
fluctuations.

Alternatively one can say that for low dimensional systems
the leading-order approximation gives qualitatively wrong results.
This happens, for example, when one studies the finite temperature
2+1 dimensional Nambu-Jona-Lasinio \cite{Shovkovy1} or related
(see \cite{MacKenzie,Ichinose}) models which possess a continuous
global symmetry. The leading-order calculation indicates that the
symmetry remains broken for a range of non-zero temperatures.
This result seems to contradict the predictions of the
Coleman-Mermin-Wagner-Hohenberg theorem \cite{Coleman}.
The argument is well known. The infrared region of the system is
dominated by the zero Matsubara mode of the boson field.
The boson field effectively resides in a 1+1 dimensional space. Thus,
by appealing to the above mentioned theorem, one concludes that
the symmetry breaking is forbidden. However, to show this explicitly
one has to calculate the next-to-leading order correction.
In the quasi-2D case \cite{Ichinose} the situation is not as dramatic
because the symmetry breaking transition is not forbidden,
but nonetheless the next-to-leading order correction changes the
results drastically.

Therefore one needs theoretical methods which allow one to study models
beyond the mean-field approximation. This may be most simply achieved
by adapting the methods developed in quantum field theory e.g. the
Coleman-Weinberg method \cite{EWeinberg1}. The main goal of the
article is to show how the approach based on the Coleman-Weinberg
effective potential may be used in the theory of superconductivity.

It is interesting to note that, as in the case of the well known
phenomenon of dynamical symmetry breaking which was originally
discovered in condensed matter theory, one observes that
the Coleman-Weinberg method is in some sense predated by the
Thouless approach to superconductivity \cite{Thouless}.

There is, however, a crucial difference between the
Coleman-Weinberg and Thouless methods.
In the Thouless approach, the appearance of an imaginary part in
the effective potential which occurs below the mean-field
transition temperature, $T_{c}^{MF}$, is considered as the signature of
the onset of superconductivity. This is the so called Thouless
criterion of superconductivity which is, in fact, equivalent to the
criterion given by the BCS theory. Contrary to this, a deeper analysis
of the Coleman-Weinberg effective potential \cite{EWeinberg2}
(see also \cite{Miransky,Peregoudov}) has shown that the appearance
of an imaginary part in the effective potential does not
necessarily imply a symmetry breaking, e.g. superconducting
transition. It only indicates a failure in the approximations used
to derive the potential. It should prove useful and interesting to
apply these concepts to the theory of superconductivity.

In Section 2 we present the model and derive the
tree-potential along with the Coleman-Weinberg one-loop
correction. The explanation of how the BCS mean-field results are
related to the tree-potential allows one to understand easily why
the one-loop correction is complex at the point of interest. The
problem of complexity is known to appear in quantum field theories
with spontaneous symmetry breaking at tree-level, and to correct it,
one should replace the one loop correction by the so-called modified
effective potential \cite{EWeinberg2}.  In Section 3 we approximate
the solution to the full gap equation (including the fluctuations) for
the case $T_{c} \ltwid T_{c}^{MF}$ \cite{LQSh} and thus derive a
corresponding modified effective potential.  The solution of the full
gap equation for the case $T_{c} \ll T_{c}^{MF}$ is obtained in Section 4
and our conclusions are given in Section 5.

\section{Model and Formalism}
\setcounter{equation}{0}

We study the following Hamiltonian density
\begin{equation}
H = -\psi_{\sigma}^{\dagger}(x)
\left[ \frac{\nabla_{\perp}^{2}}{2 m_{\perp}} +
\frac{1}{m_{z} d^{2}} \cos(i d \nabla_{z}) + \mu \right] \psi_{\sigma}(x)
- V \psi_{\uparrow}^{\dagger}(x) \psi_{\downarrow}^{\dagger}(x)
    \psi_{\downarrow}(x) \psi_{\uparrow}(x),     \label{Hamiltonian}
\end{equation}
which describes a layered quasi-2D superconductor with coherent
interlayer tunneling. Here $x \equiv  \tau, \mbox{\bf r}_{\perp}, r_{z}$
(with imaginary time $\tau$ and $\mbox{\bf r}_{\perp}$ being a 2D vector);
$\psi_{\sigma}(x)$ is a fermion field,
$\sigma = \uparrow, \downarrow$ is the spin variable;
$m_{\perp}$ is the effective carrier mass in the planes
(for example CuO$_2$ planes);
$m_{z}$ is an effective mass in the $z$-direction;
$d$ is the interlayer distance;
$V$ is an effective local attraction constant;
$\mu$ is a chemical potential which fixes the carrier density $n_{f}$;
and we take $\hbar = k_{B} = 1$.

The relevance of the Hamiltonian (\ref{Hamiltonian}) to the description
of HTSC is discussed in \cite{LQSh,QSh} and the references therein.

From a field theoretical point of view the closest model to the one
considered here was studied in \cite{Ichinose}. This model is the
quasi-(2+1) dimensional Nambu-Jona-Lasinio model which has a
very similar and rather rich phase diagram \cite{Ichinose}.
However, it is not our purpose here to discuss the whole phase diagram
of our model here as was done in \cite{LQSh,QSh} and we will restrict
ourselves to the field theoretical aspects of the transition to the
phase with broken symmetry. There is however an important difference
between the model considered in \cite{Ichinose} and the nonrelativistic
model here.  Due to the presence in (\ref{Hamiltonian}) of the chemical
potential $\mu$ the density of particles in this model can be varied and
may be large while in \cite{Ichinose} the fermion density is not fixed.
It is also known that in the presence of the Fermi surface the symmetry
breaking transition may happen for arbitrarily small attraction, while in
the Nambu-Jona-Lasinio model the attraction must be stronger then some
critical value.

Note that only the case $T_{c} \ll T_{c}^{MF}$  was considered
in \cite{Ichinose} (recall that $T_{c}^{MF}$ is the
mean-field transition temperature and $T_{c}$ is the critical
temperature of the superconducting transition). This is related
to the assumption that the anisotropy of the quasi-2D system
is sufficiently high. However, in the present work, one cannot make this
assumption since, if the anisotropy of the system is not very large,
one expects that $T_{c} \ltwid T_{c}^{MF}$ for high carrier density
\cite{Kats,Efetov}. Thus we have to study here both the possibilities
$T_{c} \ltwid T_{c}^{MF}$ \cite{LQSh} and $T_{c} \ll T_{c}^{MF}$.

The standard Hubbard-Stratonovich method was used to study the Hamiltonian
(\ref{Hamiltonian}) (see for example the review \cite{Kleinert}). In this
method the statistical sum $Z(v,\mu,T)$ ($v$ is the volume of the system)
is formally rewritten as a functional
integral over the auxiliary Hubbard-Stratonovich fields
$\Phi$ and $\Phi^{\ast}$
\begin{equation}
Z(v,\mu,T) = \int {\cal D} \Phi {\cal D} \Phi^{\ast}
\exp[-\beta \Omega(v, \mu, T, \Phi(x), \Phi^{\ast}(x))], \label{statsum}
\end{equation}
where
\begin{equation}
\beta \Omega(v, \mu, T, \Phi(x), \Phi^{\ast}(x)) =
\frac{1}{V} \int_{0}^{\beta} d \tau \int d
\mbox{\bf r} |\Phi(x)|^{2} - \mbox{Tr Ln} G^{-1}[\Phi(x),\Phi^{\ast}(x)]
                                    \label{Effective.Action}
\end{equation}
is the one-loop effective action. Here the auxilliary fields are
given at the classical level by the equations of motion $\Phi = V
\psi_{\downarrow} \psi_{\uparrow}$ and $\Phi^{\ast} = V
\psi_{\uparrow}^{\dagger} \psi_{\downarrow}^{\dagger}$.

The action (\ref{Effective.Action})
is expressed in terms of the Green function $G$ which has
in the Nambu representation (see e.g. \cite{Schrieffer})
the following operator form
\begin{equation}
G^{-1}[\Phi(x),\Phi^{\ast}(x)] =
- \hat{I} \partial_{\tau} +
\tau_{3} \left[\frac{\nabla_{\perp}^{2}}{2 m_{\perp}} +
\frac{1}{m_{z} d^{2}} \cos(i d \nabla_{z}) + \mu \right]
+ \tau_{+} \Phi(x) + \tau_{-} \Phi^{\ast}(x),
                          \label{Green.initial}
\end{equation}
where $\tau_{3}$, $\tau_{\pm} = (\tau_{1} \pm i \tau_{2})/2$ are
Pauli matrices.
Although the representation (\ref{statsum}), (\ref{Effective.Action}) is
exact, in practical calculations it is necessary to restrict ourselves to
some approximation. For our purposes the most convenient approximation is
the Coleman-Weinberg \cite{EWeinberg1} (see also \cite{Miransky,Peregoudov})
effective potential in the one-loop approximation.  The exact expression
(\ref{statsum}) is replaced by
\begin{equation}
Z(v,\mu,T, |\Phi|^{2}) =
\exp[-\beta \Omega_{pot}(v, \mu, T, |\Phi|^{2})],
\label{statsum.Gaussian}
\end{equation}
where the effective thermodynamical
potential
\begin{equation}
\Omega_{pot}(v, \mu, T, |\Phi|^{2}) \simeq
\Omega_{pot}^{\mbox{\tiny MF}}(v, \mu, T, |\Phi|^{2}) +
\Omega^{(1)}(v, \mu, T, |\Phi|^{2})
                                    \label{Effective.Potential}
\end{equation}
is expressed through the mean-field ``tree-potential''
\begin{equation}
\Omega_{pot}^{\mbox{\tiny MF}}(v, \mu, T, |\Phi|^{2}) =
\left. \Omega(v, \mu, T, \Phi(x), \Phi^{\ast}(x))
\right|_{\Phi, \Phi^{\ast} = \: \mbox{\tiny const}}
                                    \label{Tree.Potential}
\end{equation}
which may be evaluated explicitly to give
\begin{equation}
\Omega _{pot}^{\mbox{\tiny MF}}(v, \mu, T, |\Phi|^2) = {v} \left[
\frac{|\Phi|^2}{V} - \int \frac{d \mbox{\bf k}}{(2 \pi)^{3}}
\left\{2T
\ln\cosh{\frac{\sqrt{\xi^{2}(\mbox{\bf k})     + |\Phi|^2}}{2 T}} -
\xi(\mbox{\bf k}) \right\} \right],
                     \label{Tree.Potential.final}
\end{equation}
where
\begin{equation}
\xi(\mbox{\bf k}) = \mbox{\bf k}_{\perp}^{2}/2m_{\perp}
- w \cos k_{z} d- \mu, \qquad w=(m_z d^2)^{-1}
\label{sp.energy}
\end{equation}
and through the one-loop (quantum) correction
\begin{equation}
\Omega^{(1)}(v, \mu, T, |\Phi|^{2}) =
\frac{T}{2} \mbox{Tr Ln} \Gamma^{-1}(\tau, \mbox{\bf r})
|_{\Phi, \Phi^{\ast} = \: \mbox{\tiny const}}
                                    \label{Quantum.Potential.exact}
\end{equation}
with
\begin{equation}
\Gamma^{-1}(\tau, \mbox{\bf r}) =
\left(
\begin{array}{cc}
\frac{\ds \beta \delta^{2} \Omega}
{\ds \delta \Phi^{\ast} (\tau, \mbox{\bf r}) \delta \Phi(0,\mbox{\bf 0})}
&
\frac{\ds \beta \delta^{2} \Omega}
{\ds \delta \Phi(\tau, \mbox{\bf r}) \delta \Phi(0,\mbox{\bf 0})} \\
\frac{\ds \beta \delta^{2} \Omega}
{\ds \delta \Phi^{\ast}(\tau, \mbox{\bf r}) \delta \Phi^{\ast}(0,\mbox{\bf 0})}
&
\frac{\ds \beta \delta^{2} \Omega}
{\ds \delta \Phi(\tau, \mbox{\bf r}) \delta \Phi^{\ast}(0,\mbox{\bf 0})}
\end{array}
\right).                             \label{Green.fluct}
\end{equation}
Here and above the Tr and Ln are understood in the functional sense.

Using the ``tree-potential'' (\ref{Tree.Potential.final}) one can
easily reproduce the results of the BCS theory. Indeed the condition of
minimization
\begin{equation}
\left. \frac{\partial \Omega_{pot}^{\mbox{\tiny MF}}(v, \mu, T, |\Phi|^{2})}
{\partial |\Phi|^{2}} \right|_{|\Phi| =
\Phi_{min}^{(0)}}  = 0 \label{gap.def}
\end{equation}
gives the standard BCS gap equation
\begin{equation}
\frac{1}{V} = \int \frac{d \mbox{\bf k}}{(2 \pi)^{3}}
\frac{1}{2 \sqrt{\xi^{2}(\mbox{\bf k})+ (\Phi_{min}^{(0)})^{2}}}
\tanh{\frac{ \sqrt{\xi^{2}(\mbox{\bf k)} + (\Phi_{min}^{(0)})^{2}}}{2 T}},
                             \label{gap}
\end{equation}
where  $\xi(\mbox{\bf k)}$ is given by (\ref{sp.energy}).

In addition to the gap equation (\ref{gap.def}) there is the
condition
\begin{equation}
-\frac{1}{v} \frac{\partial
\Omega_{pot}^{\mbox{\tiny MF}}(v, \mu, T, |\Phi|^{2})}{\partial \mu}
= n_{f},                     \label{number.def}
\end{equation}
which fixes the carrier density $n_{f}$.
Since we have a quasi-2D system with a quadratic dispersion law in
the planes the Fermi energy $\epsilon_F$ is given by
\begin{equation}
\epsilon_F = \frac{\pi n_F d}{m_\perp}.
\end{equation}

In the standard BCS theory the carrier density is so high that
the feedback effect from the formation of the superconducting order
parameter $\Phi$ on the chemical potential $\mu$ is negligible
and the equation (\ref{number.def}) has a trivial solution
$\mu = \epsilon_{F}$. Although for small carrier densities
the equation (\ref{number.def}) becomes very important
(see for the review \cite{LSh.review}), we consider here the high
density limit only.

In this limit one can approximately split the quantum correction
(\ref{Quantum.Potential.exact}) written using the matrix
(\ref{Green.fluct}) into the following sum
\begin{equation}
\Omega^{(1)}(v, \mu, T, |\Phi|^{2}) =
\frac{T}{2} \mbox{Tr Ln} \Gamma_{-}^{-1}(\tau, \mbox{\bf r}) +
\frac{T}{2} \mbox{Tr Ln} \Gamma_{+}^{-1}(\tau, \mbox{\bf r}) \, ,
                                    \label{Quantum.Potential}
\end{equation}
where the Green functions $\Gamma_{\pm}$ are
\begin{eqnarray}
\Gamma_{\pm}^{-1} (\tau, \mbox{\bf r})
=  && \frac{1}{V} \delta(\tau) \delta(\mbox{\bf r}) +
\mbox{tr} [G(\tau, \mbox{\bf r}) \tau_{+} G(-\tau, -\mbox{\bf r}) \tau_{-}]
\nonumber              \\
&& \pm \left.
\mbox{tr} [G(\tau, \mbox{\bf r}) \tau_{-} G(-\tau, -\mbox{\bf r}) \tau_{-}]
\right|_{\Phi = \Phi^{\ast} = \: \mbox{\tiny const}}.
                \label{Gamma(pm).def}
\end{eqnarray}
This splitting is valid if these Green functions are even functions
of momenta and frequency in the momentum representation.
It was shown in \cite{Popov} that this assumption is justified when one
uses these functions in the derivative approximation, i.e. when they
are expanded for small momenta and energy.
This is the case considered in what follows.

In the weak coupling limit for local attraction between carriers
it is appropriate to replace the attraction constant $V$ by the
two-particle bound state energy in vacuum \cite{Miyake,GGL},
\begin{equation}
\varepsilon_b = - 2W \exp \left( - \frac{4 \pi d}{m_\perp V} \right).
\label{bound.energy}
\end{equation}
Here $W$ is the bandwidth in the plane and the limit $V \rightarrow 0$,
$W \rightarrow \infty$ is to be understood. This replacement enables one to
regularize the ultraviolet divergences which are present in the
four-Fermi theory. Recall that in the case of non-local phonon attraction
they are usually removed by the introduction of a cutoff at the Debye
frequency \cite{Schrieffer}. This simplifies our condensed matter problem
since one always has a natural scale for regularization.

Let us recall how the shape of the potential
$\Omega_{pot}^{\mbox{\tiny MF}}(|\Phi|^{2})$
and the solution of Eq.(\ref{gap.def}) depend on $T$.
Above the mean-field critical temperature $T_{c}^{MF}$ the equation
(\ref{gap.def}) only has the trivial solution $\Phi = \Phi^{\ast} =0$, i.e.
the ``tree-potential'' (\ref{Tree.Potential}) is everywhere convex i.e.\\
$\partial^{2} \Omega_{pot}^{\mbox{\tiny MF}}/ \partial \Phi \partial
\Phi^{\ast} > 0$, with a minimum at $\Phi = \Phi^{\ast} = 0$.

The mean-field temperature, $T_{c}^{MF}$, is defined by the
following equation
\begin{equation}
\left. \frac{\partial
\Omega_{pot}^{\mbox{\tiny MF}}(v, \mu, T_{c}^{MF}, |\Phi|^{2})}
{\partial |\Phi|^{2}} \right|_{\Phi = \Phi^{\ast} = 0} = 0,
                                   \label{TcMF.def}
\end{equation}
which results in the standard BCS equation
\begin{equation}
\frac{1}{V} = \int \frac{d \mbox{\bf k}}{(2 \pi)^{3}}
\frac{1}{2 \xi(\mbox{\bf k})}
\tanh{\frac{\xi(\mbox{\bf k)}}{2 T_{c}^{MF}}}.             \label{TcMF}
\end{equation}
Taking into account the renormalization by (\ref{bound.energy})
described above and using $(m_z d^2)^{-1} \ll T_{c}^{MF}$ one obtains
from (\ref{TcMF}) that
\begin{equation}
T_{c}^{MF} = \frac{\gamma}{\pi} \sqrt{2 |\varepsilon_{b}| \epsilon_{F}},
               \label{TcMF.sol}
\end{equation}
where $\ln \gamma = 0.577$ is the Euler constant.

Thus at the temperature $T_{c}^{MF}$ the curvature
of the ``tree-potential'' around the point $\Phi = \Phi^{\ast} = 0$
changes from convex to non-convex, i.e. \\
$\partial^{2} \Omega_{pot}^{\mbox{\tiny MF}}(T_c^{MF})/
\partial \Phi \partial \Phi^{\ast}|_{\Phi = \Phi^{\ast} =0} = 0$.

Finally, for $T < T_{c}^{MF}$ the equation (\ref{gap.def})
has a nontrivial degenerate solution which, for $T \ltwid T_{c}^{MF}$,
satisfies
\begin{equation}
|\Phi(T)|^2 = \Phi_{min}^{(0)}(T)^2 =
\frac{8 \pi^{2} (T_{c}^{MF})^{2}}{7 \zeta(3)}
\left( 1 - \frac{T}{T_{c}^{MF}} \right),
                        \label{gap.depen}
\end{equation}
where $\zeta(3)$ is the zeta function.
This in turn means that $\Omega_{pot}^{\mbox{\tiny MF}}(|\Phi|^{2})$
has a degenerate minimum for which $|\Phi(T)| = \Phi_{min}^{(0)}(T)
\neq 0$, while the point $\Phi = \Phi^{\ast} =0$ becomes a maximum so
that the ``tree-potential'' is non-convex around this point,
$\partial^{2} \Omega_{pot}^{\mbox{\tiny MF}}/
\partial \Phi \partial \Phi^{\ast}|_{\Phi = \Phi^{\ast} =0} < 0$.

A gap equation which already includes the fluctuation correction
(\ref{Quantum.Potential}) has the same form as Eq.(\ref{gap.def}),
but with the ``tree-potential'' replaced by the  full potential
(\ref{Effective.Potential}):
\begin{equation}
\frac{\partial
\Omega_{pot}(v, \mu, T, |\Phi|^{2})}
{\partial |\Phi|^{2}}  = 0 .
                                   \label{gap.full.def}
\end{equation}
Correspondingly, the critical temperature, $T_{c}$ is defined by
\begin{equation}
\left. \frac{\partial
\Omega_{pot}(v, \mu, T_{c}, |\Phi|^{2})}
{\partial |\Phi|^{2}} \right|_{\Phi = \Phi^{\ast} = 0} = 0.
                                   \label{Tc.def}
\end{equation}
Since $T_{c} < T_{c}^{MF}$ it follows that the ``tree-potential'' is
concave around the point $\Phi = \Phi^{\ast} = 0$ at $T_c$. Nonetheless
the full potential $\Omega_{pot}$ must be convex i.e.
$
\partial^{2} \Omega_{pot}/ \partial \Phi \partial \Phi^{\ast}|_
{\Phi = \Phi^{\ast} =0} > 0
$
around the point $\Phi = \Phi^{\ast} = 0$ for
$T > T_{c}$ since otherwise $T_{c}$ would not be the critical
temperature. It follows from this that for $T_{c} < T < T_{c}^{MF}$
the quantum correction (\ref{Quantum.Potential}) has to transform
the maxima at the point $\Phi = \Phi^{\ast} = 0$ to the minima of
the full potential $\Omega_{pot}(|\Phi|^{2})$.

Unfortunately, the one-loop correction $\Omega^{(1)}$ given by
(\ref{Quantum.Potential}) is ill-defined (complex) at the point of
interest. The potential develops a complex part as the temperature
drops below $T = T_{c}^{MF}$ and in the earlier development of the BCS
theory \cite{Thouless} this was considered as the sign of
superconductivity (the Thouless criterion of superconductivity).
However, this situation is rather standard in quantum field theory if
one considers the class of theories with tree-level symmetry breaking
and an extensive literature exists \cite{EWeinberg2,Miransky,Stevenson,Sher}.
Thus from a field theoretical point of view, the appearance of the
imaginary part of the effective potential is only an indication that
the one-loop approximation fails near the point $\Phi = \Phi^{\ast} = 0$
and does not mean the symmetry is broken in the next-to-leading order
approximation. Furthermore the imaginary part is related to the
non-convexity of $\Omega_{pot}^{MF}(|\Phi|^{2})$ at this point which
makes the argument of logarithm in (\ref{Quantum.Potential}) in the
momentum space negative and this non-convexity is related with the
symmetry breaking on the tree level.  It is important to note that the
discussion in \cite{EWeinberg2,Miransky,Sher} is valid only when the
quantum correction is written in the diagonal form (\ref{Quantum.Potential})
rather than  for the nondiagonal case (\ref{Quantum.Potential.exact}).

There are many ways to circumvent the complexity of the effective
potential explained above. For example, one can use the so-called
Gaussian \cite{Stevenson} or modified \cite{EWeinberg2} effective potential
which coincides with $\Omega_{pot}$ in the region where the latter is
well-defined and is still well-defined in the region where the original
effective potential is ill-defined.

It is possible to find the modified potential directly, but we will
use here, in our opinion, a more transparent consideration which will
allow us to evaluate the one-loop correction to the gap equation not at
$\Phi=0$ where it is ill-defined but in the region where it is well-defined.
This will be done in the next section where the solution to the full gap
equation for $T_{c} \ltwid T_{c}^{MF}$ is obtained.
We will also discuss there how our corresponding modified potential relates
to that introduced in \cite{EWeinberg2}. For simplicity we may assume from
now on that $\Phi$ is real without loss of generality.

\section{The solution for $T_{c} \ltwid T_{c}^{MF}$}
\setcounter{equation}{0}

Let us assume that $T_{c} \ltwid T_{c}^{MF}$ which means the position
$\Phi_{min}^{(0)}$  of the minimum of $\Omega_{pot}^{MF}(|\Phi|^{2})$
is close to zero. At this point $\Omega_{pot}^{MF}$ is definitely
convex and the one-loop correction (\ref{Quantum.Potential}) is real and
well-defined. Thus for $T_{c} \ltwid T_{c}^{MF}$ one can approximate
Eq.(\ref{Tc.def}) by
\begin{equation}
\frac{1}{v} \left.
\frac{\partial \Omega_{pot}^{MF}(v, \epsilon_{F},T_{c}, \Phi^{2})}{\partial
\Phi^{2}}  \right|_{\Phi = 0} +
\frac{1}{v} \left.
\frac{\partial \Omega^{(1)}(v, \epsilon_{F},T_{c}, \Phi^{2})}{\partial
\Phi^{2}}  \right|_{\Phi = \Phi_{min}^{(0)}} = 0 \,,
                             \label{Bgap.approx}
\end{equation}
where the value of $\Phi_{min}^{(0)}$
at temperature $T \ltwid T_{c}^{MF}$ is given by (\ref{gap.depen}).

Therefore to solve the approximated gap equation (\ref{Bgap.approx})
one has to calculate
\begin{eqnarray}
&& \frac{1}{v} \left.
\frac{\partial \Omega^{(1)}(v, \epsilon_{F},T, \Phi^{2})}
{\partial \Phi^{2}} \right|_{\Phi = \Phi_{min}^{(0)}}
\nonumber              \\
&& = \frac{1}{2} \frac{T}{(2 \pi)^{3}}
\sum_{\pm} \sum_{n = -\infty}^{\infty} \int d \mbox{\bf K}
\Gamma_{\pm} (i \Omega_{n}, \mbox{\bf K}) \left.  \frac{\partial
\Gamma_{\pm}^{-1} (i \Omega_{n}, \mbox{\bf K})} {\partial \Phi^{2}}
\right|_{\Phi = \Phi_{min}^{(0)}}\, ,
                 \label{derivative}
\end{eqnarray}
where ${\bf K}=({\bf K}_\perp,K_z)$. Starting from (\ref{Gamma(pm).def}),
one can obtain the Green's functions as a function of $\Phi$ in
the momentum representation as
\begin{equation}
\Gamma_{\pm}^{-1}(i \Omega_{n}, \mbox{\bf K}) =
\frac{1}{V}  + \frac{T}{(2 \pi)^{3}} \sum_{l = -\infty}^{\infty} \int d
\mbox{\bf k} \frac{[i \omega_{l} - \xi_{-}] [i \omega_{l} + i \Omega_{n} +
\xi_{+}] \pm \Phi^{2}} {[\omega_{l}^{2} + \xi_{-}^{2} + \Phi^{2}]
[(\omega_{l} + \Omega_{n})^{2} + \xi_{+}^{2} + \Phi^{2}]},
                           \label{Gamma(pm)}
\end{equation}
where we have introduced the abbreviations
$\xi_{\pm} = \xi(\mbox{\bf k} \pm \mbox{\bf K}/2)$ and
$\Omega_{n} = 2 \pi n T$ , $\omega_{l} = \pi (2l + 1) T$ are odd and even
Matsubara frequencies, respectively.

Since $\Gamma^{-1}_-(0,{\bf 0}) = 0$ is nothing but the BCS gap equation,
the solution to the equation is the BCS value $\Phi_{min}^{(0)}$. In
other words at $\Phi=\Phi_{min}^{(0)}$, $\Gamma^{-1}_-(0,\mbox{\bf K})$
has a zero at $\mbox{\bf K} = 0$. This gives rise to a pole in
$\Gamma_{-}(0, \mbox{\bf K})$ at $\mbox{\bf K} = 0$ which is
the only singular term in (\ref{derivative}) for $\Phi = \Phi_{min}^{(0)}$.
\begin{equation}
\frac{1}{v} \left.
\frac{\partial \Omega^{(1)}(v, \epsilon_{F},T, \Phi^{2})}{\partial \Phi^{2}}
\right|_{\Phi = \Phi_{min}^{(0)}}  \simeq  \frac{1}{2}
\frac{T}{(2 \pi)^{3}} \int d \mbox{\bf K} \left.
\Gamma_{-} (0, \mbox{\bf K}) \right|_{\Phi=\Phi_{min}^{(0)}}
\left.  \frac{\partial \Gamma_{-}^{-1} (0, \mbox{\bf K})}{\partial \Phi^{2}}
\right|_{\Phi = \Phi_{min}^{(0)}}
                     \label{derivative.approx}
\end{equation}

In order to perform the calculations analytically we use the
high-temperature derivative
expansion for the Green functions $\Gamma^{-1}_{\pm}$ \cite{LQSh},
\begin{equation}
\Gamma_{\pm}^{-1} (\Phi^{2}; 0, \mbox{\bf K}) =
\frac{m_{\perp}}{2 \pi d}
\left[ \ln \left( \frac{T_c}{T_c^{MF}} \right) + a\mbox{\bf K}_{\perp}^{2}
+ b (1 - \cos{K_z d}) +
c (2 \Phi^{2} \pm \Phi^{2}) \right] \, ,
                    \label{Gamma(pm).der}
\end{equation}
where
\begin{equation}
a = \frac{7 \zeta(3)}{(4 \pi)^2} \frac{\epsilon_{F}}{m_{\perp} T^2},
\qquad b = \frac{7 \zeta(3)}{(4 \pi)^2} \frac{w^2}{T^2}, \qquad
c = \frac{7 \zeta(3)}{8 \pi^{2} T^{2}}.
                                      \label{abc}
\end{equation}

One can check that $\Gamma_{-}^{-1}(0, \mbox{\bf 0})$ becomes negative
for $\Phi < \Phi_{min}^{(0)}$ and this results in the appearance of the
imaginary part of the effective potential described in the previous section.

Strictly speaking the one-loop correction (\ref{Quantum.Potential.exact})
should be also real in the region \\
$\Phi_{conv} < \Phi < \Phi_{min}^{(0)}$,
where $\Phi_{conv}$ is the point where the convexity of
$\Omega_{pot}^{MF}(|\Phi|^{2})$ changes sign. However, due to the
approximation implicit in the factorization needed to
obtain (\ref{Quantum.Potential}) one observes that the one-loop correction
is complex for $\Phi < \Phi_{min}^{(0)}$.  This is the reason why we used
in the approximated gap equation (\ref{Bgap.approx}) the point
$\Phi_{min}^{(0)}$ for the calculation of the value of the effective
potential rather than the point $\Phi_{conv}$.

Substituting (\ref{Gamma(pm).der}) with $\Phi_{min}^{(0)}$
given by (\ref{gap.depen}) into
(\ref{derivative.approx}), one arrives at the following approximation
\begin{equation}
\frac{1}{v} \left.
\frac{\partial \Omega^{(1)}(v, \epsilon_{F},T, \Phi^{2})}{\partial \Phi^{2}}
\right|_{\Phi = \Phi_{min}^{(0)}}  \simeq
\nonumber              \\
\frac{T}{(2 \pi)^{3}}
\int d \mbox{\bf K} \frac{c}
{a \mbox{\bf K}_{\perp}^{2} + b [1 - \cos{K_z d}]}.
                                     \label{derivative.approx.1}
\end{equation}
One can see that Eq.(\ref{derivative.approx.1}) has no infrared
divergencies due to the presence of the third
direction ($b \neq 0$). In two dimensions it would be infrared
divergent as required by the 2D theorems \cite{Coleman}.
This equation also has an artificial ultraviolet
divergence as a result of the replacement
of the Green's function $\Gamma$ by its derivative approximation.
Thus one should introduce a rather natural ultraviolet cutoff
$(\mbox{\bf K}_{\perp}^{max})^{2} = 2 m_{\perp} \Phi(T=0) =
2 m_{\perp} \sqrt{2 |\varepsilon_{b}| \epsilon_{F}}$
and integrate over the momentum $\mbox{\bf K}$ to
obtain the expression
\begin{equation}
\frac{1}{v} \left.
\frac{\partial \Omega^{(1)}(v, \epsilon_{F},T, \Phi^{2})}{\partial \Phi^{2}}
\right|_{\Phi = \Phi_{min}^{(0)}}  \simeq
\frac{m_{\perp}}{2 \pi d} \frac{T}{2 \epsilon_{F}} |\ln \kappa| \, ,
                                     \label{correction.final}
\end{equation}
where
\begin{equation}
\kappa = \frac{1}{4 \sqrt{2}} \frac{w^2}{\epsilon_{F}^{2}}
\sqrt{\frac{\epsilon_{F}}{|\varepsilon_{b}|}}.
                                       \label{kappa}
\end{equation}
Substituting (\ref{correction.final}) into (\ref{Bgap.approx})
one obtains the final transcendental equation for $T_{c}$
\begin{equation}
\ln \frac{T_c}{T_{c}^{MF}} + \frac{T_c}{2 \epsilon_{F}}
|\ln \kappa| = 0 \, ,
              \label{Tc.eq}
\end{equation}
which may be rewritten in the following more convenient form
\begin{equation}
T_{c} = 2 \epsilon_{F} \frac{|\ln(T_{c}/T_{c}^{MF})|}{|\ln \kappa|}.
                       \label{Tc}
\end{equation}
One can see $T_{c}$ goes to zero as $m_{z} \to \infty$ ($w \to 0$)
as it must \cite{Coleman}.

Strictly speaking the equation (\ref{Tc}) is only valid when
$T_c \ltwid T_{c}^{MF}$. Therefore one may expand the
logarithm in equation (\ref{Tc}) to obtain the equation
\begin{equation}
T_{c} = T_{c}^{MF} \left(1 +
\frac{T_{c}^{MF} |\ln\kappa|}{2 \epsilon_{F}} \right)^{-1}.
            \label{Tc.near}
\end{equation}

As stated above one can also understand the
approximation used in (\ref{Bgap.approx}) in terms of the modified
effective potential defined in \cite{EWeinberg1}. The modified effective
potential in \cite{EWeinberg1} is defined as the minimum value for $\Omega$
given a homogeneous state where $|\Phi|^2$ is uniform. The real part of this
modified potential has the following form
\begin{equation}
\tilde \Omega^{(1)}(v, \mu, T, |\Phi|^{2}) =
\frac{T}{2 (2 \pi)^{3}} \sum_{\pm} \sum_{n = -\infty}^{\infty}
\int_{\cal D} d \mbox{\bf K} \ln
\Gamma_{\pm}^{-1} (i \Omega_{n}, \mbox{\bf K})
\, ,
             \label{Quantum.Potential.modified}
\end{equation}
where the area $\cal D$ of integration in the momentum space
includes only positive modes. One can see that
(\ref{Quantum.Potential.modified}) indeed coincides with
(\ref{Quantum.Potential}) when $\Omega^{(1)}$ is well-defined.
Furthermore the modified potential (\ref{Quantum.Potential.modified})
leads to the gap equation (\ref{Bgap.approx}) which was considered
above as the approximated one.

In the region $\Phi < \Phi_{min}^{(0)}$ the modified effective potential
considered above differs from the traditional effective potential,
$\Omega_{eff}(\Phi)$, which is defined as the minimum value for $\Omega$
such that the space average of $\Phi(x)$ is given by $\Phi$. It can be shown
that the conventional effective potential is in fact the convex envelope of
the modified effective potential and is real and convex everywhere. However
for $\Phi < \Phi_{min}^{(0)}$ it describes an inhomogeneous mixed state
where the value of $\Phi(x)$ is not uniform in space. One can readily
understand that the modified and not the original potential is
relevant for the superconducting state.

There is, however, the difference between our interpretation of the
modified potential and that of \cite{EWeinberg2}. In \cite{EWeinberg2}
the homogeneous state described by (\ref{Quantum.Potential.modified})
is considered as decaying and the rate of the decay is related to
negative modes of (\ref{Quantum.Potential}) which are not included in
(\ref{Quantum.Potential.modified}). It is physically obvious that
there is no real decay of the homogeneous superconducting state with
$\Phi < \Phi_{min}^{(0)}$ for $T < T_{c}$ although we have not been able
to prove this rigorously. The absence of decay is in agreement with the
interpretation of \cite{Stevenson} although it should be stressed that
the modified potential discussed here is not identical to the
Gaussian effective potential in \cite{Stevenson}.

\section{The solution for $T_{c} \ll T_{c}^{MF  }$}
\setcounter{equation}{0}

In this section we find the approximate solution of
Eq.(\ref{Tc.def}) for the case $T_{c} \ll T_{c}^{MF}$,
i.e. when the anisotropy is large ($w/ |\varepsilon_{b}| \ll 1$).

Following \cite{Ichinose} we plan to expand about the minimum
of the mean-field potential (\ref{Tree.Potential}) i.e. about the
 point $\Phi_{min}^{(0)}$, which is determined by Eq.(\ref{gap.def}).
Including the fluctuations shifts the minimum at $\Phi_{min}^{(0)}$
to $\Phi = \Phi_{min}^{(0)} + \Phi^{(1)}$. One may therefore approximate
(\ref{gap.full.def}), expanding about the mean-field minimum, as
\begin{equation}
\frac{\partial \Omega_{pot}(\Phi^{2})} {\partial \Phi^{2}} \simeq
\frac{\partial \Omega_{pot}(\sigma)} {\partial \sigma} +
\frac{\partial^{2} \Omega_{pot}(\sigma)} {(\partial \sigma)^{2}}
2 \Phi_{min}^{(0)} \Phi^{(1)} = 0,                      \label{Bgap}
\end{equation}
where we have introduced the short-hand notation
$\sigma \equiv (\Phi_{min}^{(0)})^{2}$. This equation can be simplified
using the ``tree'' gap equation (\ref{gap.def}) and one arrives at
\begin{equation}
\Phi^{(1)} = - \frac{\partial \Omega^{(1)}(\sigma)}{\partial \sigma}
\left\{ 2 \Phi_{min}^{(0)} \left[
\frac{\partial^{2} \Omega_{pot}^{\mbox{\tiny MF}}(\sigma)}
{(\partial \sigma)^{2}} +
\frac{\partial^{2} \Omega^{(1)}(\sigma)} {(\partial \sigma)^{2}}
\right] \right\}^{-1}.
                                           \label{Bgap.final}
\end{equation}
The temperature $T_{c}$ is defined by the condition
\begin{equation}
\Phi(T_{c}) = \Phi_{min}^{(0)} (T_{c}) + \Phi^{(1)} (T_{c}) =0.
                                  \label{critical.temp}
\end{equation}
Substituting $\Phi_{min}^{(0)}(T_c)$ from (\ref{gap.def}) and $\Phi^{(1)}$
from (\ref{Bgap.final}) into (\ref{critical.temp}) one arrives at
the equation for $T_{c}$ in the same approximation as in \cite{Ichinose}.
It can be easily shown that the limiting behaviour $T_{c} \to 0$
when $m_{z} \to \infty$ is related to the behaviour of
$\partial \Omega^{1}(\sigma)/ \partial \sigma$ (see below).
For this reason for large $m_{z}$ one can replace
$\Phi_{min}^{(0)}(T_{c}, m_{z})$ and
$\Omega_{pot}^{\mbox{\tiny MF}}(T_{c}, m_{z})$ by their 2D values
at zero temperature, $\Phi^{(0)}(T = 0, m_{z} \to \infty)$ and
$\Omega_{pot}^{\mbox{\tiny MF}}(T = 0, m_{z} \to \infty)$, namely
\begin{equation}
\Phi_{min}^{(0)}(T_{c},m_z) \simeq \sqrt{2 |\varepsilon_{b}| \epsilon_{F}}
                         \label{Phi.0}
\end{equation}
and
\begin{equation}
\frac{\partial^{2} \Omega_{pot}^{\mbox{\tiny MF}}(\sigma)}
{(\partial \sigma)^{2}} \simeq
v \frac{m_{\perp}}{4 \pi d}
\frac{1}{|\varepsilon_{b}| (2\epsilon_{F} + |\varepsilon_{b}|)},
                                     \label{second.der}
\end{equation}
where instead of $V$ we used again the two-body bound state energy
$\varepsilon_{b}$ (\ref{bound.energy}). One can check that the second
derivative of $\Omega^{(1)}$ is negligible relative to the corresponding
derivative of $\Omega_{pot}^{\mbox{\tiny MF}}$.
Thus one need only calculate the first derivative of $\Omega^{(1)}$.

The potential $\Omega^{(1)}$ defined by (\ref{Quantum.Potential})
contains two terms: one involving the Green function $\Gamma_{+}$ and
one involving $\Gamma_{-}$. In the low temperature region and
for $\Phi_{min}^{(0)}$ defined by (\ref{gap.depen})
the $\Gamma_{+}$-excitations are massive, while the
$\Gamma_{-}$-excitations are massless (the Goldstone mode) \cite{Popov}.
Thus the gapped $\Gamma_{+}$-excitations are irrelevant and one can
safely consider only the $\Gamma_{-}$ part to obtain
\begin{equation}
\frac{\partial \Omega^{(1)}(\sigma)}{\partial \sigma} =
\frac{1}{2} \frac{v T}{(2 \pi)^{3}}
\sum_{n = - \infty}^{\infty} \int d \mbox{\bf K} \Gamma_{-}(i \Omega_{n},
\mbox{\bf K}) \frac{\partial \Gamma_{-}(i \Omega_{n}, \mbox{\bf K})}
{\partial \sigma},
                                      \label{first.der.def}
\end{equation}
where the Green function $\Gamma_{-}(i \Omega_{n}, \mbox{\bf K})$
has been given in (\ref{Gamma(pm)}).

Since one knows that $\Gamma_{-}(0,\mbox{\bf 0}) = 0$ for
$\Phi = \Phi_{min}^{(0)}$, as discussed after (\ref{Gamma(pm)}),
the main contribution to the integral is concentrated near zero and it can be
evaluated approximately as
\begin{equation}
\frac{\partial \Omega^{(1)}(\sigma)}{\partial \sigma} \simeq
\frac{1}{2} \frac{v T}{(2 \pi)^{3}} \sum_{n = - \infty}^{\infty}
\int d \mbox{\bf K} \Gamma_{-}(i \Omega_{n}, \mbox{\bf K})
\left. \frac{\partial \Gamma_{-}(0, \mbox{\bf 0})}{\partial \sigma}
\right|_{m_{z} \to \infty, T=0},               \label{first.der}
\end{equation}
where we have left $m_{z}$ finite and $T \neq 0$ only where necessary.

Differentiating (\ref{Gamma(pm)}) one arrives at
\begin{eqnarray}
&& \left. \frac{\partial \Gamma_{-}(0, \mbox{\bf 0})}{\partial \sigma}
\right|_{m_{z} \to \infty, T=0} =
T \sum_{l = -\infty}^{\infty} \int \frac{d \mbox{\bf k}}{(2 \pi)^{3}}
\left. \frac{1}{[\omega_{l}^{2} + \xi^{2}(\mbox{\bf k}) + \sigma]^{2}}
\right|_{m_{z} \to \infty, T=0}
\nonumber               \\
&& =
\frac{1}{4 \pi} \frac{m_{\perp}}{d} \frac{1}{(\Phi_{min}^{(0)})^{2}}.
                                               \label{derivative.zero}
\end{eqnarray}

Thus one needs only the expression for $\Gamma_{-}(i \Omega_{n},
\mbox{\bf K})$. In the low temperature limit this Green function for small
$\Omega_{n}$ and $\mbox{\bf K}$ (the derivative approximation) takes
the following form
\begin{equation}
\Gamma_{-}^{-1}(i \Omega_{n}, \mbox{\bf K}) =
a^{\prime} \mbox{\bf K}_{\perp}^{2} +b^{\prime} K_{z}^{2}
+ d^{\prime} \Omega_{n}^{2}, \qquad
\Omega_{n}, \frac{\mbox{\bf K}_{\perp}^2}{m_{\perp}},
\frac{K_{z}^{2}}{m_{z}} \ll \Phi_{min}^{(0)} (T = 0)
                                      \label{deriv.expan}
\end{equation}
with
\begin{equation}
a^{\prime} = \frac{\epsilon_{F}}{8 \pi d (\Phi^{(0)}_{min})^{2}}, \qquad
b^{\prime} = \frac{1}{32 \pi d} \frac{m_{\perp}}{m_{z}}
    \frac{w}{\epsilon_{F}^{2}}, \qquad
d^{\prime} = \frac{m_{\perp}}{8 \pi d (\Phi^{(0)}_{min})^{2}}.
                                  \label{abd}
\end{equation}
To obtain (\ref{deriv.expan}) one has to remove the ultraviolet divergences
from (\ref{Gamma(pm)}) by applying the procedure of regularization via
the two-body bound state energy described in Section 2.  We note that in the
limit $T \to 0$ one can expand $\Gamma_-^{-1}$ over $\Omega_{n}$ without
performing the analytical continuation to real frequency $\Omega$.  Note
also that after this analytic continuation is performed the pole of
$\Gamma_{R -}^{-1} (\Omega, \mbox{\bf K}_{\perp}, K_{z}=0) = 0$
gives the correct dispersion relation for the Bogolyubov mode
\begin{equation}
\Omega = \sqrt{\epsilon_{F}/m_{\perp}} K_{\perp} = v_{F}/\sqrt{2} K_{\perp}.
\end{equation}
Using the simple expression (\ref{deriv.expan}) makes it possible
to calculate analytically $\partial \Omega^{(1)}/ \partial \sigma$
given by (\ref{first.der}).

One can easily see that, in the 2D case when $b^\prime = 0$, the term
in the integral (\ref{first.der}) with $\Omega_{n} = 0$ is infrared
divergent. This in turn implies that the gap equation (\ref{critical.temp})
only has the trivial solution $T_{c} \equiv 0$ which again demonstrates
the absence of long range order in 2D \cite{Coleman}.

If one uses the expansion (\ref{deriv.expan}), the equation
(\ref{first.der}) also has an artificial ultraviolet divergence.
This is simply related to the use of the derivative expansion
(\ref{deriv.expan}) which is only valid if $\Omega_{n}$ and
$\mbox{\bf K}$ are small. One may use the same natural energy cutoff
as in the previous case, namely the BCS gap $\Phi_{min}^{(0)} (T=0)$, to
eliminate this divergence.

In the limit of large $m_{z}$ the only relevant term in the sum
(\ref{first.der}) is $\Omega_{n} = 0$ since this is the divergent term
in the 2D limit. Using the abovementioned cutoff  one arrives at
\begin{equation}
\frac{1}{2} \frac{v T}{(2 \pi)^{3}} \sum_{n = - \infty}^{\infty}
\int d \mbox{\bf K} \Gamma_{-}(i \Omega_{n}, \mbox{\bf K})
\simeq v T \frac{(\Phi_{min}^{(0)})^{2}}{\epsilon_{F}}
|\ln \delta |,
                                      \label{integral}
\end{equation}
where
\begin{equation}
\delta =
\frac{\pi^{2}}{4 \sqrt{2}}
\sqrt{\frac{|\varepsilon_{b}|}{ \epsilon_{F} }}
\frac{w^{2}}{\epsilon_{F}^{2}}.
                                            \label{delta}
\end{equation}
Substituting (\ref{integral}) into (\ref{first.der})
and then (\ref{first.der}) into (\ref{Bgap.final}) yields
\begin{equation}
\Phi^{(1)}(T) \simeq - \frac{T}{\epsilon_F} |\ln \delta| \frac{|\epsilon_b|
(2 \epsilon_F + |\epsilon_b|)}{2 \Phi^{(0)}_{min}}
\label{eq:phi1}
\end{equation}
Finally substituting (\ref{eq:phi1}) into (\ref{critical.temp}) one
obtains the expression for the critical temperature as
\begin{equation}
T_{c} \simeq \frac{2 \epsilon_{F}}{|\ln \delta|}.
                              \label{Tc.low}
\end{equation}
This result is valid only for $T_{c} \ll T_{c}^{MF}$
i.e. only when $\delta \ll 1$.

One can see that the expressions (\ref{Tc}) and (\ref{Tc.low}) both
display logarithmic singularities in the limit $m_z \rightarrow \infty$.
Also in both cases the leading order dependence on the Fermi energy
$\epsilon_F$ is linear. This is in contrast to the square root
mean-field BCS dependence (\ref{TcMF.sol}) and resembles the
experimentally observed dependence \cite{Uemura}. Thus our two rather
different approximations give qualitatively the same results.

\section{Conclusion}

To summarize, the appearance of an imaginary part in the one-loop
effective potential does not signal the onset of superconductivity.
Instead it reflects a well-known failure of the one-loop approximation.
By reference to the modified effective potential one may derive a new
approximation to the critical temperature in the limit $T_c \ltwid T_c^{MF}$.
One may also derive an approximation to the critical temperature in the
"zero-temperature" limit.  Unlike the mean-field critical temperature both
these approximate critical temperatures tend to zero in the 2D limit in
agreement with the Coleman-Mermin-Wagner-Hohenberg theorem \cite{Coleman}.
In addition they both display a roughly linear dependence on the Fermi
energy in agreement with experiment.

\section*{Acknowledgments}
We gratefully acknowledge fruitful discussions with N.J.~Davidson,
E.V.~Gorbar, V.P.~Gusynin, V.M.~Loktev, V.A.~Miransky and I.A.~Shovkovy.
One of us (S.G.Sh) is grateful to the members of the Department of Physics
of the University of Pretoria for hospitality. We acknowledge the financial
support of the Foundation for Research Development, Pretoria.

\end{document}